\documentclass[prb,preprint]{revtex4}

\usepackage{amssymb,amsmath} 
\usepackage{graphicx}

\def\eq#1{{Eq.~(\ref{#1})}}

\begin{document}

\title{A simple derivation of the electromagnetic field of an arbitrarily moving charge}

\author{Hamsa Padmanabhan}
\email{hamsa.padmanabhan@gmail.com}

\altaffiliation[ ]{Address for correspondence: 8, Akashganga, IUCAA, Pune University Campus, Ganeshkhind, Pune 411007.} 
\affiliation{Fergusson College (Pune University), Pune 411004, India}


\begin{abstract}
The expression for the electromagnetic field of a charge moving along an arbitrary trajectory is obtained in a direct, elegant, and Lorentz invariant manner without resorting to more complicated procedures such as differentiation of the Li\'enard-Wiechert potentials. The derivation uses arguments based on Lorentz invariance and a physically transparent expression originally due to J.\ J.\ Thomson for the field of a charge that experiences an impulsive acceleration.
\end{abstract}

\maketitle

\section{Introduction}

Determining the electromagnetic field of a charge moving along an arbitrary trajectory is a standard textbook problem in electrodynamics. Conventionally, the electromagnetic field of a charge moving along an arbitrary trajectory is obtained by solving the wave equation to obtain the Li\'enard-Wiechert potentials and then calculating the required derivatives of the potentials with respect to position and time.\cite{landau} The Li\'enard-Wiechert potentials are found to be expressible as functions of the retarded time, which is an implicit function of the position at which the field is being evaluated and the trajectory of the charge. This implicit dependence makes differentiating the potentials an elaborate procedure, involving complicated algebra and obscuring the physical significance of the final result.

We present here an alternative derivation of the electromagnetic field of an arbitrarily moving charge, using manifestly Lorentz invariant, four-dimensional notation, without ever obtaining the Li\'enard-Wiechert potentials or differentiating any expression with respect to the retarded time. 

The key idea is that we can derive the electromagnetic fields of a charge in the Lorentz frame at which the charge was instantaneously at rest at the retarded time. In principle, making a Lorentz transformation will give the fields in the original frame.\cite{hamsakvpy} In practice, the second step turns out to be as laborious as the standard derivation and the algebra is complicated and unilluminating. In addition to the fact that Lorentz transformations mix up $\mathbf{E}$ and $\mathbf{B}$, we need to express all the original 3-vectors in terms of the corresponding 3-vectors in the new frame. In doing so, we cannot use the standard Lorentz transformation equations which are valid when the velocity is along the x-axis (say), but must instead use the more complicated Lorentz formula for the velocity along an arbitrary direction.  Therefore, we adopt a different, manifestly Lorentz invariant approach to tackle the problem, but in the same spirit.

The essential physical ingredients in this derivation are the following: The electromagnetic field at the observation point can depend only on the relative position, the velocity, and the acceleration of the charge, all evaluated at the retarded time, but not on further derivatives of the trajectory. The electromagnetic field of a charge that experiences an impulsive acceleration, that is, an arbitrary but finite acceleration for a short time interval, can be obtained from physical arguments. This derivation is originally due to J.\ J.\ Thomson.\cite{thomson} The generalization of the Thomson formula to a charge moving with any nonzero (and possibly relativistic) velocity is possible by using arguments based on Lorentz invariance. 

The plan of the paper is as follows. In Sec.~\ref{sec:two} we construct the most general expression for the electromagnetic field tensor $F^{ab}$ in terms of the variables in the problem. This general expression contains three arbitrary functions of the Lorentz scalars, which need to be determined using further information. In Sec.~\ref{sec:three} we fix these functions by using an independent physical argument based on the Thomson formula and thus determine $F^{ab}$ completely. Our resulting expression for the electric field matches that obtained by the standard procedure of differentiating the Li\'enard-Wiechert potentials. We summarize our results in a brief concluding section.

\section{The general expression for $F^{ab}$ due to a charge} \label{sec:two}

Consider a charge moving along an arbitrary trajectory $z^i(\tau)$ whose electromagnetic field $F^{ab}(x^i)$ at the observation point $x^i$ is to be evaluated.
We will use four-dimensional notation with signature --, +,+,+; the Latin letters $i$, $j$, $k$\ldots run through $0, 1, 2, 3$ and the Greek letters $\alpha$, $\beta$, $\gamma$, denote the spatial components 1, 2, 3. We shall use units in which $c = 1$. In the conventional approach the current four-vector $J^i(x^i)$ is first obtained from the trajectory of the charge. Then, using the relation $\square A^i = -4\pi J^i $, the four-potential $A^i$ (the Li\'enard-Wiechert potentials) is calculated. The electromagnetic field tensor $F^{ij}$ is then found by the relation $F^{ij} = \partial^i A^j - \partial^j A^i$. The definition makes clear that $F^{ij}$ is antisymmetric, that is, $F^{ij} = -F^{ji}$. The $F^{0\alpha}$ terms give the components of the electric field, and the $F^{\alpha \beta}$ terms lead to the components of the magnetic field. This procedure of differentiating the Li\'enard-Wiechert potentials with respect to $x^i$ is tedious because these potentials are expressible only as functions of the retarded proper time $\tau_{\rm ret}$ which is an implicit function of $x^i$ and $z^i(\tau)$.

In our alternative approach to this problem we begin with a simple physical idea.
We note that the electromagnetic field at the observation point $x^i$ may depend only on the relative position $R^i = x^i - z^i(\tau)$, the velocity $u^i$, and the acceleration $a^i = du^i/d\tau$ of the charge, all evaluated at the retarded time $\tau_{\rm ret}$, but not on further derivatives of the trajectory. This result arises from the following: 

(a) Because electromagnetic signals propagate at the speed of light, the field at $x^i$ is determined by the state of the source at an earlier position $z^i(\tau_{\rm ret})$ which is related to $x^i$ by a null line; that is, by the condition $R_iR^i = 0$. Of the two roots to this equation, we choose the retarded (causal) solution that satisfies the condition $R^0 > 0$. This condition determines the retarded time $\tau_{\rm ret}$. 

(b) Translational invariance implies that the field depends only on the relative position $R^i$ of the charge with respect to the observation point (evaluated at the retarded time), and not on the absolute positions of the source or the observation point separately. 

(c) Because $\square A^i \sim J^i$, $F^{ik}$ satisfies $\square F^{ik} \sim \partial^iJ^k - \partial^kJ^i$. Because $J^i$ is at most linear in the velocity of the charge, $\partial^iJ^k$ is at most linear in the acceleration, and no further derivatives of the trajectory can occur in the solution $F^{ik}$. 

Therefore, $F^{ij}$ is a second rank antisymmetric tensor which is built from $R^i$, $u^i$, and $a^i$. At this stage it is convenient to introduce 
the Lorentz invariant scalar $\ell = R_iu^i$  which in the rest frame of the charge reduces to
\begin{equation}
\ell=R_iu^i=-R^0=-|\mathbf{R}|\equiv -R,
\end{equation}
where $(R^0)^2= |\mathbf{R}|^2$ because of the condition $R_iR^i=0$ and $R^0>0$ for the retarded solution. For simplicity, we will also define a four-vector $n^i$ through the relation $R^i \equiv -\ell(n^i + u^i)$. It is easy to see that $n_ku^k = 0$, and $n_kn^k = 1$. The components of $n^i$ are
\begin{equation}
n^i = \Big(-\frac{R}{\ell} - \gamma,-\frac{\mathbf{R}}{\ell} - \gamma \mathbf{v} \Big),
\end{equation} 
which reduces in the rest frame of the charge to the unit spatial vector pointing from the charge to the field point:
$n^i = (0,\mathbf{1})$. We will trade off the $R^i$ dependence of $F^{ij}$ for the $n^i$ dependence and treat $F^{ij}$ as a function of $n^i$, $u^i$, and $a^i$ (instead of $R^i$, $u^i$, and $a^i$). Hence
the most general form of $F^{ij}$ can be written as 
\begin{equation}
\label{fijgen1}
F^{ij} = f_1 a^{[i} u^{j]} + f_2 n^{[i} a^{j]} + f_3 n^{[i} u^{j]},
\end{equation} 
where the square brackets denote antisymmetrization (that is, $a^{[i} u^{j]} = a^i u^j - a^j u^i$) and the functions $f_1$, $f_2$, $f_3$ can depend on the scalars that can be constructed from $n^i$, $u^i$, and $a^i$. Because $u_iu^i = -1$, $a_iu^i = 0$, $n_iu^i = 0$, and $n_in^i = 1$, the only non-trivial scalars which are available to us are $g \equiv n_ia^i$, $R_iu^i = \ell$, and $a_ia^i$. Because $F^{ij}$ is at most linear in $a^i$, it cannot depend on $a_ia^i$, and hence $f_1$, $f_2$, and $f_3$ can depend only on the remaining two scalars $g$ and $\ell$. Further, the condition of linearity in $a^i$ also requires that $f_1$ and $f_2$ cannot depend on $g$ (because these functions are multiplied by a term linear in $a^i$) and can depend only on $\ell$. Hence, the most general expression for $F^{ij}$ in terms of $n^i$, $u^i$, and $a^i$ becomes:
\begin{equation}
\label{fijgen}
F^{ij} = f_1(\ell) a^{[i} u^{j]} + f_2(\ell) n^{[i} a^{j]} + f_3(g,\ell) n^{[i} u^{j]},
\end{equation}
with $f_1$, $f_2$, and $f_3$ arbitrary functions of the scalars which are as yet undetermined.

We comment that in constructing the most general second rank antisymmetric tensor out of $n^i$, $u^i$, and $a^i$, we can also include the three terms of the form $\epsilon_{ijkl}a^{[i} u^{j]}$ involving the totally antisymmetric tensor $\epsilon_{ijkl}$. The reason for omitting terms of this type in Eq.~\eqref{fijgen} is as follows. When $\epsilon_{ijkl}$ multiplies an antisymmetric (true) tensor $A^{ij}$, for example, we obtain the dual tensor $*A^{ij}$ which is not a true tensor, but a pseudotensor; that is, it transforms differently from true tensors under reflections of the co-ordinate system. We know that the electromagnetic field tensor $F^{ij}$ is a true tensor. Hence, we cannot have pseudotensorial terms such as $\epsilon_{ijkl}a^{[i} u^{j]}$ in the general expression for $F^{ij}$.

Our next task is to determine $f_1$, $f_2$, and $f_3$, which will completely solve the problem. We proceed by considering the two four vectors $E^i$ and $B^i$ defined as
\begin{equation}
\label{ebfij}
E^i = u_jF^{ij},
\quad
B^i = \frac{1}{2}\epsilon^{ijkl}u_jF_{kl}.
\end{equation}
The vectors $E^i$ and $B^i$ contain the same amount of information as $F^{ij}$ as can be seen by the explicit expression for the latter in terms of the former:
\begin{equation}
\label{fijalternate}
F^{ij} = u^i E^j - E^i u^j - \epsilon^{ij}\smallskip _{kl}\ u^k B^l,
\end{equation} 
which can be easily verified by direct substitution of \eq{fijalternate} into Eq.~(\ref{ebfij}) and the use of the identities $u_jE^j = 0$ and $u_j B^j = 0$. Incidentally, these identities, which follow from the definitions in \eq{ebfij} and the antisymmetry of $F^{ij}$ and $\epsilon^{ijkl}$, also show that $E^i$ and $B^i$ are both orthogonal to $u^i$, and hence, in a given reference frame, they contain only three independent components.
We substitute \eq{fijgen} into \eq{ebfij} and find 
\begin{equation}
\label{new1}
E^i=-f_1(\ell)a^i - f_3(g,\ell)n^i, \quad B^i= f_2(\ell)\epsilon^{ijkl}u_jn_ka_l.
\end{equation}
If we know the explicit forms of $E^i$ and $B^i$ in terms of the variables in the problem, we can determine the functions $f_1$, $f_2$, and $f_3$ by direct comparison. 

The four vectors $E^i$ and $B^i$ have direct physical interpretations and represent the electric and magnetic fields in the instantaneous rest frame of the charge with four velocity $u^i$.
In the instantaneous rest frame of the charge in which $u^j = (1,\mathbf{0})$ only the component $u_0$ contributes and $E^i = u_0F^{i0} = (0,F^{0\alpha})$, because $F^{ij}$ is antisymmetric. Hence, the spatial components of $E^i = u_jF^{ij}$ correctly represent the components of the electric field in the instantaneous rest frame of the charge. Similarly, in the instantaneous rest frame, only the component $u_0$ contributes to $B^i$. Because 
$\epsilon^{ijkl}$ is completely antisymmetric, the time component of $B^i$ vanishes in this frame. We see that the spatial components of $B^i$ are given by $F^{\alpha\, \beta}$ where $\alpha, \beta = 1$, 2, or 3. Hence the spatial components of $B^i$ lead to the correct values of the magnetic field components in the rest frame.

We now will obtain explicit expressions for the (three-dimensional) 
electric and magnetic field vectors in the rest frame by using an independent physical argument. From these explicit expressions we will construct two four-vectors, $\mathcal{E}^i$ and $\mathcal{B}^i$, such that the spatial components of $\mathcal{E}^i$ and $\mathcal{B}^i$ are equivalent to the electric and magnetic three-vectors in the rest frame and the time components of $\mathcal{E}^i$ and $\mathcal{B}^i$ vanish in this frame. If we use the fact that $\mathcal{E}^i$ and $\mathcal{B}^i$ are generally covariant four-dimensional quantities and are identical in the rest frame with the original $E^{i}$ and $B^{i}$ (by construction), we can equate the general expressions, $E^{i}$ to $\mathcal{E}^i$ and $B^{i}$ to $\mathcal{B}^i$, and thereby determine the functions $f_1$, $f_2$, and $f_3$.

\section{Electromagnetic fields in the instantaneous rest frame of the charge} \label{sec:three}

The physical argument to determine the electric and magnetic field vectors is based on the physically transparent expression for the electromagnetic field due to J.\ J.\ Thomson.\cite{thomson} 
This expression is usually given for a charged particle moving with non-relativistic velocities and becomes exact in the limit of $v\to0$. Its derivation is reproduced and discussed in standard texts.\cite{tap1,berkeley} For convenience we give a brief derivation in the Appendix. The analysis considers a charge that is moving with a uniform (non-relativistic) velocity, experiences a finite acceleration $\mathbf a$ for an infinitesimal time interval $\Delta t$, and then continues to move with uniform velocity. The expression for the electromagnetic field under these conditions can be expressed as:
\begin{equation}\label{field1}
\mathbf{E} = \mathbf{E}_{\rm coul} + \mathbf{E}_{\rm rad},\quad \mathbf{B} = \mathbf{\hat n}\times \mathbf{E}_{\rm rad},
\end{equation}
where
\begin{equation}
\mathbf{E}_{\rm coul} = \frac{q}{R^2}\mathbf{\hat n}
\end{equation} 
is the Coulomb field of a charge at rest a distance $R$ from the observation point with $\mathbf{\hat n}=\mathbf{R}/R$ the unit vector in the radial direction. Also 
\begin{equation} \label{thomaccn}
\mathbf{E}_{\rm rad} = \frac{q}{R}\mathbf{\hat n}\times(\mathbf{\hat n} \times \mathbf a) 
\end{equation}
is the radiation field of the charge, transverse to the radial direction, due to the acceleration $\mathbf a$ of the charge. We can thus rewrite \eq{field1} as:
\begin{align}\label{3}
\mathbf{E} &= \frac{q}{R^2}\mathbf{\hat n} + \frac{q}{R}[\mathbf{\hat n}(\mathbf{\hat n} \cdot \mathbf{a}) - \mathbf a ] \\
\label{4}
\mathbf{B} & = - \frac{q}{R} \left[ \mathbf{\hat n}\times \mathbf{a}\right].
\end{align} 
In conventional units the radiation field terms (in Eqs.~(\ref{thomaccn}), \eqref{3}, and (\ref{4})) will each contain a $c^2$ factor in the denominator. 

We now show that the electric and magnetic fields in Eqs.~(\ref{3}) and (\ref{4}) also represent the electromagnetic field of a charge in arbitrary motion, in the instantaneous rest frame at the retarded time. Although this fact is intuitively obvious, we give a formal argument to make our derivation complete. Let us compare the fields of two charges $C$ and $C_1$ that have two different trajectories.
The charge $C$ is the one of interest and moves along an arbitrary trajectory. We make a Lorentz transformation to the instantaneous rest frame (at the retarded time) of $C$. In this frame the charge $C$ is at rest, but it has an instantaneous acceleration $\mathbf{a}$. The charge $C_1$ is chosen to have the same acceleration and position as charge $C$ for a small time interval $\Delta t$ around the retarded time and moves with a uniform velocity later on. The position (at the origin), velocity (equal to zero), and acceleration (equal to $\mathbf{a}$) of $C$ and $C_1$ are identical at the retarded time, and because the electromagnetic field depends only on these quantities, both charges will lead to the same fields at the observation point. The trajectories of the two charges are quite different after a time $\Delta t$ (with $C$ moving along some arbitrary curve while $C_1$ moves with a uniform velocity), but the expression for the radiation field, \eq{thomaccn}, cannot depend on the future trajectory of charge $C$. Therefore, we have obtained the important result that in the instantaneous rest frame of the charge $C$, Eqs.~(\ref{3}) and (\ref{4}) give the correct form of the electromagnetic field.

Equations~(\ref{3}) and (\ref{4}) are three-dimensional expressions for the electric and magnetic fields. We use these expressions to now construct four-vectors, $\mathcal{E}^i$ and $\mathcal{B}^i$, such that their spatial components give the above three-vectors in the rest frame, and their time components vanish in that frame. We begin by noting that $R$ in \eq{thomaccn} is the value of the Lorentz scalar $-R_iu^i = -\ell$ in the rest frame. Also, in the rest frame we have $n^i = (0,\mathbf{1})$ and $a_i = (0, \mathbf{a})$. 
We use these results to generalize \eq{3} to four-dimensions as
\begin{equation}\label{6}
\mathcal{E}^i = \frac{q}{\ell^2} n^i + \frac{q}{\ell}[ a^i - n^i (n_ka^k) ].
\end{equation}
The structure of this four vector is unique and easy to understand. We would have expected it to be a linear combination of $a^i$, $n^i$, and $u^i$ with coefficients depending on the scalars built out of them. But a term linear in $u^i$ cannot occur because $u^i$ in the rest frame has only the time component but $a^i$ and $n^i$ have only space components. So their linear combination cannot vanish in the rest frame (and give $\mathcal{E}^0=0$) if there is a term with $u^i$. The coefficients of $a^i$ and $n^i$ are uniquely fixed by the form of \eq{3}. Also, note that $\mathcal{E}^i$ contains one term which goes as $1/\ell^2$ and one term which goes as $1/\ell$, which come from the three-dimensional Coulomb field and radiation field respectively.

To generalize \eq{4} we note the appearance of the cross product. In three dimensions a cross product is represented by making use of the completely antisymmetric tensor $\epsilon^{\alpha \beta \gamma}$. We thus need a tensor that goes over to $\epsilon^{\alpha \beta \gamma}$ in the instantaneous rest frame. Consider the tensor $u_i\epsilon^{ijkl}$. In the rest frame, as mentioned before, the contribution is only from $u_0$. Because $\epsilon^{ijkl}$ is antisymmetric, all the components of $u_i\epsilon^{ijkl}$ when $j$, $k$, or $l$ is zero, vanish in the rest frame. Hence, the non-zero components of the tensor $u_i\epsilon^{ijkl}$ are identical to those of
$-\epsilon^{\alpha \beta \gamma}$ in the rest frame. Hence, $u_i\epsilon^{ijkl}$ is the tensor we need and \eq{4} can thus be generalized as 
\begin{equation}\label{7}
\mathcal{B}^i = \frac{q}{\ell}\epsilon^{ijkl}u_jn_ka_l.
\end{equation} 
We see that in the rest frame, the time component of Eq.~\eqref{7} vanishes and the spatial components give the magnetic field.

All that remains is to equate the explicit forms for $\mathcal{E}^i$ and $\mathcal{B}^i$ in Eqs.~\eqref{6} and \eqref{7} to the general expressions for $E^{i}$ and $B^{i}$
in \eq{new1} involving the undetermined functions $f_1$, $f_2$, and $f_3$ and determine the latter.
The equality $\mathcal{E}^i=E^i$ gives
\begin{equation}
-f_1(\ell )a^i - f_3(g,\ell )n^i = \frac{q}{\ell ^2}n^i + \frac{q}{\ell }a^i - \frac{q}{\ell }n^i(n_ka^k).
\label{this}
\end{equation} 
Equation~\eqref{this} determines the functions $f_1$ and $f_3$ to be (recall that $g=n_ka^k$)
\begin{equation}
f_1(\ell ) = -\frac{q}{\ell }, \quad f_3(g,\ell ) = \frac{q}{\ell }(n_ka^k) - \frac{q}{\ell ^2}.
\end{equation}
Similarly, the equality $\mathcal{B}^i=B^i$ gives
\begin{equation}
f_2(\ell )\epsilon^{ijkl}u_jn_ka_l = \frac{q}{\ell }\epsilon^{ijkl}u_jn_ka_l,
\end{equation} 
and fixes $f_2$ to be
\begin{equation}
f_2(\ell ) = \frac{q}{\ell }.
\end{equation} 
Thus we obtain the final expression for $F^{ij}$:
\begin{equation}\label{fijfinal}
F^{ij} = \frac{q}{\ell ^2}u^{[i} n^{j]} + \frac{q}{\ell }\left[ n^{[i} a^{j]} - a^{[i} u^{j]} + (n_ka^k) n^{[i} u^{j]} \right].
\end{equation}
Equation~\eqref{fijfinal} completely solves the problem. 
The electromagnetic field tensor $F^{ij}$ naturally splits into a ``Coulomb" term and a ``radiation" term.

As an aside, we note that there is an alternative and somewhat shorter way of arriving at \eq{fijfinal}. 
We substitute the four-dimensional generalized fields derived from the Thomson expression, namely $\mathcal{E}^i$ and $\mathcal{B}^i$, for $E^i$ and $B^i$ respectively into \eq{fijalternate} and obtain the explicit expression for $F^{ij}$ without first arriving at the expression in \eq{fijgen}. If we substitute Eqs.~(\ref{6}) and (\ref{7}) into \eq{fijalternate}, we obtain 
\begin{equation}\label{fijalternate1}
F^{ij} = \frac{q}{\ell ^2}u^{[i} n^{j]} - \frac{q}{\ell }a^{[i} u^{j]} + \frac{q}{\ell }(n_ka^k)n^{[i} u^{j]} -
\frac{q}{\ell }\epsilon^{ij}\smallskip _{kl}\epsilon^{lpqr} u^k n_q a_r u_p.
\end{equation} 
To evaluate the expression $\epsilon^{ij}\,_{kl}\epsilon^{lpqr} u^k n_q a_r u_p$ we use the identity
\begin{equation}\label{epsilon}
\epsilon_{ijkl}\epsilon^{lpqr} = -[\delta^p_i(\delta^r_j\delta^q_k - \delta^q_j\delta^r_k) - \delta^r_i(\delta^p_j\delta^q_k - \delta^p_k\delta^q_j) + \delta^q_i(\delta^p_j\delta^r_k - \delta^r_j\delta^p_k)].
\end{equation} 
We lower the indices $i$ and $j$ in \eq{fijalternate1} and then use \eq{epsilon} to obtain the expression for $F_{ij}$:
\begin{equation}
F_{ij} = \frac{q}{\ell ^2}u_{[i} n_{j]} - \frac{q}{\ell }a_{[i} u_{j]} + \frac{q}{\ell }(n_ka^k)n_{[i} u_{j]} + \frac{q}{\ell }n_{[i} a_{j]}.
\end{equation}
Raising the indices $i$ and $j$ reproduces \eq{fijfinal}. 

The final result, \eq{fijfinal}, is known in the literature. It is obtained by integrating Maxwell's equations in a four-dimensional notation, and differentiating the resultant Li\'enard-Wiechert potentials $A^j$ with respect to $x^i$.\cite{tap1,jackson}

As a direct check of \eq{fijfinal}, we evaluate the $F^{0\alpha}$ components of \eq{fijfinal}. To do so we expand \eq{fijfinal} with $i = 0$ and $j = \alpha$ in terms of the zero and $\alpha$ components of the four-vectors $u^i$, $n^i$, and $a^i$. If we use 
\begin{align}
 n^i &= \left(-\frac{R}{\ell} - \gamma,-\frac{\mathbf{R}}{\ell} - \gamma \mathbf{v}\right) \\
 u^i & = (\gamma, \gamma \mathbf{v}) \\
 a^i & = \left(\gamma^4 \mathbf{v}\cdot\mathbf{a}, \gamma^2 \mathbf{a} + \gamma^4(\mathbf{v}\cdot\mathbf{a})\mathbf{v}\right)
\end{align}
(obtained by calculating $du^i/ d\tau$),
and the fact that in the lab frame, the value of $\ell $ is $R_iu^i = \gamma(-R + \mathbf{ v}\cdot \mathbf{R})$, it is easy to show that
\begin{equation}
F^{0\alpha} = \frac{q (1 - v^2)}{(R - \mathbf{v}\cdot\mathbf{R})^3} (\mathbf{R} -\mathbf{v}R) + \frac{q }{(R - \mathbf{v}\cdot\mathbf{R})^3} (\mathbf{R} \times ((\mathbf{R - \mathbf{v}}) \times \mathbf{a})),
\end{equation} 
which is identical to the expression for the electric field obtained by differentiating the standard Li\'enard-Wiechert potentials.\cite{landau} 

\section{Conclusion}

We have derived the general expression for the electromagnetic field of an arbitrarily moving charge in a comparatively simple manner. This approach fully exploits the nature of the dependence of the electromagnetic field tensor on the dynamical variables. The only other ingredient used is the Thomson expression, which is physically transparent and fairly straightforward to derive. The algebra is simplified considerably by the use of simple and intuitive physical arguments. Our derivation, apart from pedagogical interest, also yields the explicit expression for the electromagnetic field tensor $F^{ij}$ in terms of the position, velocity, and acceleration four-vectors. The use of four-dimensional variables and Lorentz invariant quantities does away with the need to perform any Lorentz transformations or differentiate complicated implicit functions with respect to the retarded time. 

\appendix*

\section{Derivation of the Thomson formula}

\begin{figure}[ht]
\begin{center}
\includegraphics[scale=0.5]{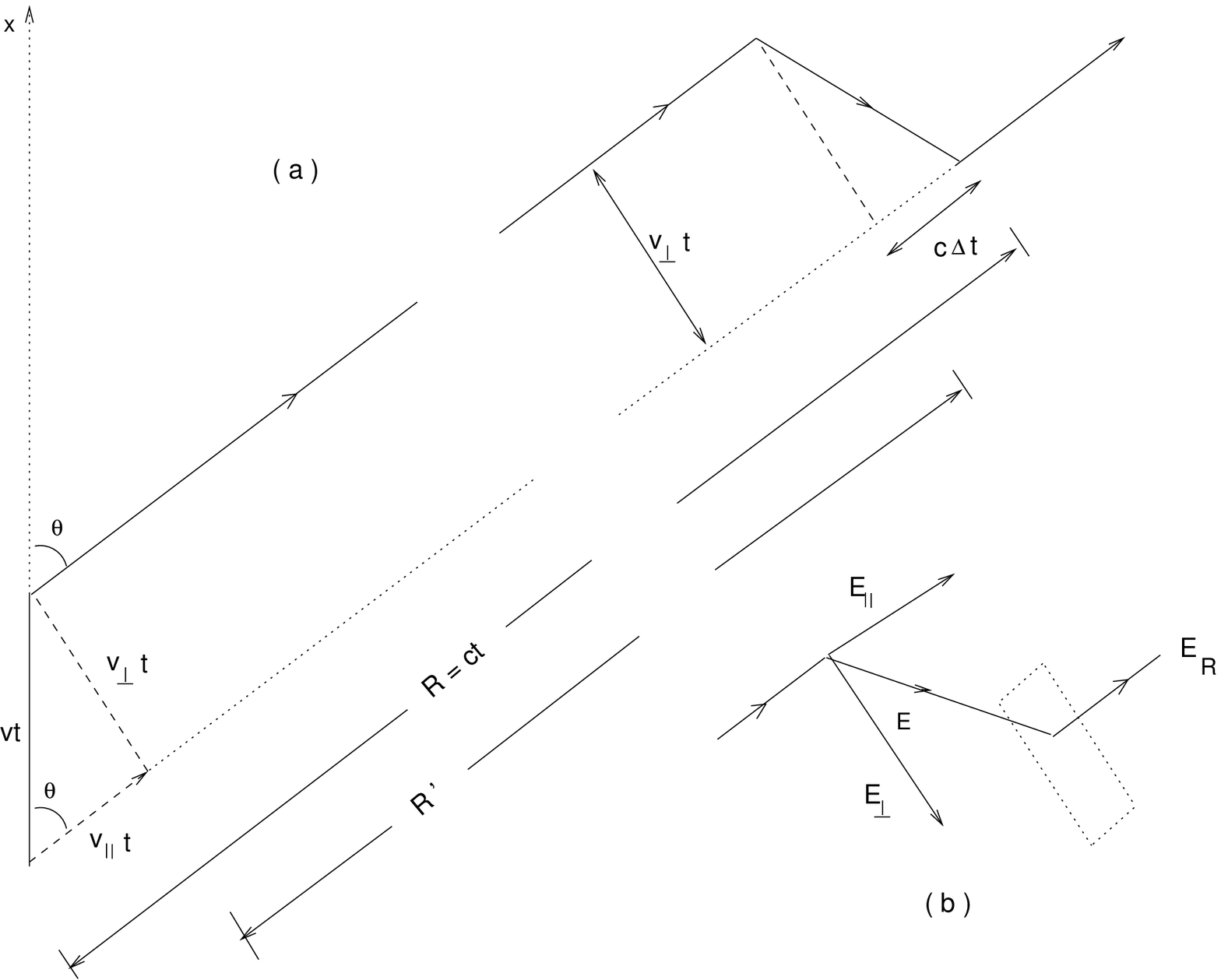}
\end{center}
\caption{(a) Electric field lines of a charge accelerated for an infinitesimal time interval $\Delta t$. (b) Gaussian pillbox used to relate $E_\parallel$ to $E_R$.}
\label{fig:simprad}
\end{figure}

We briefly derive the Thomson formula given in \eq{field1}. Consider a charged particle which is at rest until time $t = 0$, experiences an acceleration $\mathbf a$ for an infinitesimal time interval $\Delta t$, and then continues to move with uniform velocity. The electric field lines of the charge as observed at a time $t \gg \Delta t$ are shown in Fig.~\ref{fig:simprad}(a). 
We are interested in the limit $\Delta t\to 0$ with $\mathbf{a}$ finite such that the final velocity of the charge is nonrelativistic, $v=a \Delta t\ll c$. In this Appendix we have re-introduced the $c$-factor.

At distances $R > ct$ the field is that of a charge at rest: $E_R = q\mathbf{\hat n}/R^2$. At distances $R^{'} < c(t - \Delta t)$ the field is that of a charge moving with uniform velocity, which for non-relativistic velocities can be approximated to that of a charge at rest: $E_{R^{'}} = q\mathbf{\hat n^{'}}/{R^{'2}}$. Between these two limits the acceleration $\mathbf a$ of the charge produces kinks in the field lines and leads to the production of a transverse radiation field $E_{\rm rad}$ in addition to the radial Coulomb field. To find the magnitude of $E_{\rm rad}$ we resolve the velocity $\mathbf{v}$, acceleration $\mathbf{a}$, and total electric field $\mathbf{E}$ into components perpendicular and parallel to the radial unit vector $\mathbf{\hat n}$. Similarity of triangles then gives
\begin{equation}
\frac{E_\perp}{E_\parallel} = \frac{v_\perp t}{c \Delta t}.
\end{equation}
By using $v_\perp = a_\perp \Delta t$ and $t = R/c$, we get
\begin{equation}
\frac{E_\perp}{E_\parallel} = \frac{a_\perp R}{c^2}. 
\end{equation} 
Electric flux conservation applied to the Gaussian pillbox shown in Fig.~\ref{fig:simprad}(b), equates $E_\parallel$ with $E_R = q /R^2$ and thus leads to
\begin{equation}
\mathbf{E_\perp} = \frac{-q\mathbf{a_\perp}}{c^2R},
\end{equation}
which, on noting that $\mathbf{a_\perp} = \mathbf{a} - \mathbf{\hat n}(\mathbf{\hat n} \cdot \mathbf{a})$, readily gives \eq{thomaccn}.

\begin{acknowledgments} 
I thank T.\ Padmanabhan for originally suggesting this problem and guiding me toward its solution. I also thank T.\ P.\ Singh for useful discussions. Comments from Ghanashyam Date, K.\ Subramanian, Rajaram Nityananda, C.\ S.\ Unnikrishnan, Apoorva Patel, D.\ Narasimha, and Ruth Durrer on the draft of the paper are gratefully acknowledged. The initial research leading to this work was carried out while I was at TIFR, Mumbai, and I gratefully acknowledge the use of the library facilities there. It is a pleasure to acknowledge with gratitude the use of the library facilities at IUCAA, Pune. I am supported by the Kishore Vaigyanik Protsahan Yojana (KVPY) fellowship programme of DST, India.
\end{acknowledgments}


\begin{thebibliography}{99}
\bibitem{landau} See, for example, L. D. Landau and E. M. Lifshitz, \textit{The Classical Theory of Fields}
(Butterworth-Heinemann, Oxford, 1975), Chap. 8.

\bibitem{hamsakvpy} Hamsa Padmanabhan, unpublished.

\bibitem{thomson} J. J. Thomson, \textit{Electricity and Matter} (Archibald Constable, London, 1907), Chap. III. 

\bibitem{tap1} T. Padmanabhan, \textit{Theoretical Astrophysics: Astrophysical Processes} (Cambridge University Press, Cambridge, 2000), Vol. 1, Chap. 4.

\bibitem{berkeley} E. M. Purcell, \textit{Electricity and Magnetism}, The Berkeley Physics Course (Mc-Graw-Hill, New York, 2008), 2nd ed., Appendix B. Also discussed in F. S. Crawford, \textit{Waves}, The Berkeley Physics Course (Mc-Graw-Hill, New York, 1968), Chap. 7.

\bibitem{jackson} J. D. Jackson, \textit{Classical Electrodynamics} (John Wiley \& Sons, New York, 1999), 3rd ed., Chap. 14.

\end{thebibliography}
\end{document}